# Insights from Nature for Cybersecurity

Elżbieta Rzeszutko and Wojciech Mazurczyk

Warsaw University of Technology, Institute of Telecommunications

Warsaw, Poland, 00-665, Nowowiejska 15/19

Email: E.Rzeszutko@tele.pw.edu.pl, wmazurczyk@tele.pw.edu.pl

**Abstract**. The alarming rise in the quantity of malware in the last few years poses a serious challenge to the security community and requires urgent response. However, current countermeasures seem to be no longer effective. Thus, it is our belief that it is now time for researchers and security experts to turn to nature in the search for novel inspirations for defense systems. Nature has provided species with a whole range of offensive and defensive techniques, which have been developing and improving in the course of billions of years of evolution. The extremely diverse living conditions have promoted a large variation in the devised bio-security solutions. In this paper we introduce a novel PROTECTION framework in which common denominators of the encountered offensive and defensive means are proposed and presented. The bio-inspired solutions are discussed in the context of cybersecurity, where some principles have already been adopted. The deployment of the whole nature-based framework should aid the design and improvement process of modern cyber-defense systems.

**Introduction**

The growing diversity, complexity and proliferation of malware draws increasing attention from researchers and security experts, as well as ordinary Internet users. However, current countermeasures seem to be no longer effective. Few months ago, Symantec, which pioneered computer security with its antivirus software in the late 1980s, declared that their products are able to detect only about 45% of the potential threats [8]. Adding to this an almost exponential growth in newly detected malicious software, reported by security vendors in the last few years, we are currently facing a challenge that needs to be addressed urgently. The main question is: where should we seek novel ideas and inspiration that could aid the reengineering of our current defense systems. It is our belief that the answer is: in nature.

Nature has over 3.8 billion years of experience in developing solutions to the challenges facing organisms living in extremely diverse conditions. The estimated number of species is counted in millions, and each of them possesses traits ensuring survival advantage. In nature, each organism's "aim" is to fulfil two goals: *to survive* and *to reproduce*. These "goals" can be endangered by two main factors: *(i)* presence and actions of predators (or any "attacker" that could pose a threat to an organism, e.g. herbivores for plants or disease agents like viruses, bacteria or fungi); *(ii)* inability to access resources necessary for survival (like habitat, water and food). As pointed out by Sagarin and Taylor [1], *uncertainty* is a starting point when considering security in nature – as every organism tries to decrease it for itself, and increase for its potential adversaries.

In our previous work [3] we have shown, on concrete examples, that analogies between nature and cybersecurity exist. The main concepts of many current network security threats can be found in the actions of real predators. For example, an Anglerfish (Lophius Piscatorius) has eighty long filaments along the middle of its head, the most important being the longest one, terminating in a lappet, and is movable in every direction. The angler attracts other fish by means of its lure, to seize them with its enormous jaws as they approach. The same concept is currently utilized in phishing attacks, where the aim is to attract the potential victim and fool it to swallow the bait, e.g. by visiting a phony website which masquerades as a legitimate one, and fool the user to give out his credentials. Another example is the Kudzu vine which proliferates its ecosystem with astounding speed (ca. 30 cm per day). Within weeks, just like DDoS (Distributed Denial of Service) attacks for communication networks, it can literally choke all other growth, including trees and shrubs, by making it impossible to access the resources necessary for survival – light and nutrients. The same analogies can be drawn for security solutions [3].

The important observation is that in both worlds a continuous arms race takes place between offensive and defensive techniques. Due to the ever-changing scene of battle, unpredictability in cybersecurity should be considered with equal attention as it is in nature.

In literature there have already been a few attempts to transplant biologically-inspired concepts to cybersecurity. The most notable examples include: artificial immune systems [10], predator-prey association [12], malware ecology [13] or epidemic spreading [11]. Recently, the need to balance the uncertainty between the attacker and the defender has come to light in the form of e.g. various moving target techniques [1]. However, in our opinion, there is still a notable lack of a systematic review of the features, with which nature has equipped organisms to tackle the uncertainty and adapt to risks posed by even the most skilled predators and the challenging environment.

This paper introduces the concept of the bio-inspired PROTECTION framework which is intended to show the direction which the evolution of existing security solutions, and the design of future ones, should follow. By inspecting the rules and techniques that are common in nature, it was possible to identify five vital features that, in our opinion, every cybersecurity solution should employ in order to attain higher effectiveness. We illustrate how, in response to various historical challenges and threats, communication networks and security measures were (mostly unconsciously) improved to fulfil the framework's features. We also present a case study that validates our framework.

**Adaptability is the key**

In nature the answer to unpredictability is an organism's ability to *adapt* i.e. adjust its structure, behaviour and/or interactions, triggered in response to challenges or threats potentially endangering its survival. Therefore, adaptability permits *reaction* to challenges and threats as they *arise* in the environment for this particular organism. To be efficient the reaction must be made accurately and within a reasonable time. The need for adaptability propels an organism's evolution.

Sagarin et al. [1], [4] identified three most important features that have allowed organisms to survive and adapt throughout billions of years – namely: decentralization, redundancy and cooperation, and showed how they generally relate to public security and also to the ICT world. It is our belief that these features do not completely reflect the "whole picture" of adaptability. Main differences between our approach and Sagarin's et al. can be summarized as follows: *(i)* our approach attempts to determine how knowledge about nature can be practically projected onto cybersecurity while Sagarin et al. provided very general considerations on how organisms' functioning improved over time and how it relates to societal and homeland security [1], [4]; *(ii)* Sagarin et al. pointed three main features with the aid of which adaptability is realized and how it projects onto general success of a security solution.

We argue that, besides the mentioned components, two more features are vital for adaptability: *responsiveness* and *heterogeneity*, hence, there should be *five* of them which collectively form the PROTECTION framework (Fig. 1). The more components a given solution covers, the more adaptable it becomes.

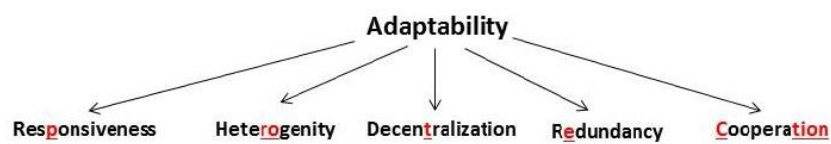

Fig. 1 PROTECTION framework for adaptability.

The features that form the PROTECTION framework with corresponding concrete examples from nature (cybersecurity examples will be provided separately in the following sections) are:

*(i)* *Responsiveness* – understood as an accurate and reactive, not proactive, behaviour once an alarming condition is arising/sensed. Prediction of challenges is seen as inaccurate and energy demanding (too costly). Biology has equipped us with a great implementation of such a responsive solution. Despite our heavily relying on the brain to govern our wellbeing, we are equipped with special neural pathways, or reflex arcs, which permit us to react instantaneously when a sensor is provoked. The electric impulses do not take the longer route from the sensor via the brain to the effector organ, but travel only to the spinal cord, where appropriate reaction is automatically triggered. Thus the reaction time is as short as possible.

*(ii)* *Heterogeneity* – the larger the number of genotypic variations in a given species, the larger is the cumulative chance for survival. Moreover, it may also involve a differentiation and specialization of roles. Heterogeneity enables evasion of the attacker's interest – the greater the diversity, the less potential benefit of putting effort into targeting a rare entity of an organism or a computer system. Certain species of moths have benefited from their natural diversification of phenotypes. The antipredatory adaptation of Peppered Moths was their speckled camouflage that came in many hues. But the Industrial Revolution altered their habitat – trees were sooted and barren of any lichens. The species as a whole survived, because natural selection favoured the darker shaded variant of moths, which reproduced freely to sustain the population.

*(iii)* *Decentralization* – may come in many forms, e.g. in the form of diverse localizations, or distributed control (increased autonomy). It capacitates prompt response to an emerging threat. This is particularly visible in non-solitary organisms, e.g. insects, forming social relationships (e.g. ants or bees) as they have developed complex means of communicating information gathered by members of the community to the collective benefit of all.

*(iv)* *Redundancy* – nature does not always strive to be optimal in its design and utilizes this feature to increase the chance for success. Redundancy is seen as the source of resilience: surplus resources allocated to a task prove essential when critical events occur. Redundancy in nature can take different forms – some lizards perform tail autotomy to evade capture, the lost limb is not crucial and can be regenerated over time; animals living in communities duplicate certain roles, thus the loss of a single member of a group is not significant.

*(v)* *Cooperation* is coined as means to push the limits of adaptability. Organisms forming symbiotic relationships with others can together overcome problems, which would otherwise be insurmountable. For example, ants protect vulnerable aphids from predators in exchange for nurturing sugary honeydew produced by the latter. Both parties benefit from this relationship.

Noting the parallelism between relationships and challenges present in nature and in the virtual world (as mentioned in [3] and in the Introduction), the proposed PROTECTION framework, in our opinion, can be conveniently projected to illustrate historical challenges, as well as current trends and developments, in communication networking and cybersecurity.

When looking into the development of communication networks one can observe that they consecutively introduce features from the PROTECTION framework to improve their overall effectiveness. An overview of the communication networks' evolution in this context is necessary as they form a sort of "canvas", or a virtual environment, for the threats, and thus they have significant influence on the cybersecurity measures.

**Communication networks and adaptability**

The evolution of communication networks follows a steady transition from centralized and homogenous systems towards a more dispersed architecture. The most notable example are the P2P (Peer-to-Peer) networks. The response to the single point of failure weakness of centralized networking and the client-server limitations was the introduction of P2P network design. The *decentralization* feature was achieved by distributing control and transport functions among the peers. At the same time *responsiveness* was ensured by amending the vulnerabilities of the previous design, i.e. resiliency increased as the network became capable of undisrupted operation even if its fragment was malfunctioning.

Decentralization was also the key to a particular service development – let us consider IP telephony. Classical, centralized VoIP (Voice over IP) solutions have been suffering issues related to NAT (Network Address Translation) and firewall traversal. On the other hand, Skype, due to its P2P design, experiences no such problems. The solution to these problems was *decentralization* and *heterogeneity*. Two roles that each Skype node can be assigned are differentiated depending on the node's network capabilities: a Super Node (SN) or an Ordinary Node (ON). Typically, SNs *cooperate* while determining user's location in a Skype network and participate during the signalling phase of any call. They also help ONs to establish and relay a call if they are behind a NAT.

Meanwhile, another P2P service – BitTorrent has taken advantage of *redundancy* and improved significantly the effectiveness of its transmissions as compared with previous P2P file sharing systems. In BitTorrent, from the network perspective, every resource is divided into many duplicate fragments and is available from multiple sources (the so called seeds and leechers). The more copies of the original resource in the network, the better, because different fragments of the same item can be downloaded simultaneously.

**The PROTECTION framework and cybersecurity**

As mentioned, the PROTECTION framework is suitable for describing the evolution of cybersecurity – the threats, as well as the countermeasures, and the aspects that should be taken into account when designing new security mechanisms. It is our belief that the more features of this framework will be incorporated into a security solution, the more adaptable, effective and consequently more secure it will become.

Looking into existing threats and countermeasures in cybersecurity, one or even a few of the framework's components can already be found. We argue that, in many cases, they have naturally and unconsciously emerged in the process of evolution. In the rest of this section we will provide examples from cybersecurity (both offensive and defensive measures) which will refer to the five introduced features of the PROTECTION framework.

Considering *decentralization*, let us take botnets as an example. First, they were built in a centralized manner, however, as they were easier to defeat this way, they adapted and evolved to P2P design. They became more difficult to detect and eliminate. At the other extreme, from the security solutions' perspective, the development of Intrusion Detection Systems (IDSs) also proved that decentralization is a good approach – localizing many "probes" (IDS sensors) increased the chance of successful threat detection.

Likewise, both offensive and defensive techniques in cybersecurity take advantage of *responsiveness*. However, many security approaches are designed to try to predict future threats rather than to effectively react to the existing, arising ones. A convenient example describing this tendency is SPIT (SPAM over Internet Telephony) which, a few years ago, was envisioned as a serious threat to future IP telephony systems, and some efforts have been made to mitigate it before it even appeared [7]. Yet, it turned out not to be a problem of the same magnitude as SPAM, and currently not many countermeasures are utilized in practice. In terms of offensive techniques, the responsiveness is somewhat more natural because the main challenge is to find a vulnerability, most likely a zero-day, and this always involves reacting and adjusting to the existing security measures. The actions are dictated by the conditions – an attacker must find a single vulnerability or a limited number of them to successfully compromise a network, while the defender must manage the whole, often complex and heterogenic security system with an unknown number of security gaps. Taking the security solutions' perspective, the reactive behaviour is visible in the Intrusion Prevention System (IPS) part of IDS/IPS system. Until recently the typical response to a new threat has been attained by creating a new malware signature, which capacitated future detection and prevention (an analogy to an immune system can be identified). However, this approach for combating malware is no longer effective because of the vast amount of malicious code with different signatures and the emergence of polymorphic and metamorphic malware that never uses the same signature more than once.

Generally, *cooperation* among cybercriminals, especially until few years ago, has been more visible and more effective than between cybersecurity experts (however now finally the situation starts to change). The black hat community has been more willing to share their knowledge, experience and "tools" than the white hat community, where everything comes at a price. An example of an attack that strongly benefits from cooperation is the well-known black hole attack on MANETs' (Mobile Ad-hoc Networks) popular routing algorithms. It relies on a malicious node deliberately injecting bogus routing information into the network to redirect the legitimate traffic to a non-existent node. The negative impact is amplified when a group of malicious nodes cooperates with each other while generating falsified routing information. From the security solutions' perspective, a recently proposed security solution that benefits from cooperation is the SocialScan [5] that enables distributed, friend-to-friend suspicious objects' scanning service with priorities governed by levels of social altruism.

*Heterogeneity* in cybersecurity is a result of increasing diversity of devices, systems and services in communication networks. Typically, to defend a network, a combination of various security systems is applied. Simultaneously, heterogeneity is exploited by cybercriminals who utilize a whole range of malware tools. There is a plethora of malware variants that continuously evolve to exploit new security vulnerabilities. According to Av-Test just during 2014 over 140 million forms of new malware have been identified in the wild. Cybercriminals exploit heterogeneity to hide the true nature of their malware – polymorphism, variable "decoy" parts of source code or instruction substitution are meant to obfuscate the true purpose of the program, evade detection, and slow down the reverse-engineering process.

In the ICT world *redundancy* is viewed as profoundly related to resilience, rather than security. When applied to the latter case, it is usually treated as an additional, dispensable cost. Moreover, applying many redundant security systems at the same location often results in ambiguous behaviour (e.g. multiple firewalls on the same device). One cybersecurity solution employing redundancy, is the honeypot, as it is basically a clone of some existing system, but established and tuned to learn the behaviour and/or the tools of cybercriminals. The recent trend in malware evolution implicates that redundancy is a valued trait. Previous worms typically utilized a single zero-day vulnerability, but more recent malware, like the infamous Stuxnet, attacked Windows systems using four unprecedented zero-day vulnerabilities. Similarly, it has been observed that certain attacks are conducted by means of double injection of the same malicious code, with the aid of two different exploits – just to make sure that the infection takes hold.

**Case study**

Advanced Persistent Threats (APTs) have been a plague affecting large organisations and governments in the past few years. These stealthy and insidious campaigns are carried out with the employment of significant financial and human resources to obtain information of potential intelligence value. Operation Aurora is one notable example of APT put to life and conducted at an unprecedented scale [15].

The events linked to Operation Aurora and the subsequent steps taken by the compromised organisations show how the elements of the PROTECTION framework can be put to life to the benefit of those involved in a security incident. In early 2010, Google was the first to admit that it had been hacked [16], alongside some US government institutions and other large companies. The attack was linked to Chinese sources, and deemed a serious threat to national security.

Instead of confining the information about the breach (as various companies often did earlier), Google took the *responsive* approach – it took the risk of spoiling its image as an "impregnable fortress", and publicly stated that it had been compromised. This in turn prompted other "smaller" victims to come to light, which triggered a whole chain of events.

First of all, Google turned to NSA (National Security Agency) to establish *cooperation* [17] in order to jointly investigate the extent of damage caused by Aurora. With Google's resources and NSA's experience it was possible to develop a tool for attack detection and response.

Actually, the created tool was *heterogenous* – it consisted of two independent modules. The first, Turmoil, was responsible for identifying the symptoms of an impending attack, while the second – Turbine, was intended for emergency countering [18]. Turbine's behavior could either be conservative – suppressing the flow of unwanted traffic or more proactive – launching a counterattack. *Decentralization* was an important success factor here – Google's vast infrastructure permitted creation of numerous, dispersed points of sensing, enabling prompt response to an emerging threat.

Lastly, *redundancy* comes into light. Although collaboration of Google and NSA was sufficient to mitigate the immediate perils linked to Aurora, it was decided that other victims and potential targets should be involved in the cyber-attack countering process [18]. Institutions of the so-called critical infrastructure were identified and regularly briefed on the current risk factors. This approach yielded an increase in security awareness among company management and limited the number of possible entry points for future attacks.

**Lessons learnt**

The main lesson from nature for cybersecurity is that no security solution is effective forever. The process of constant adaptation and evolution applies to both defensive and offensive techniques in nature and in cybersecurity. Moreover, even in nature, the best protection cannot guarantee complete, unfaltering security. For example, the most wanted defence mechanism, a replica of which every cybersecurity professional would embrace, is the human immune system. Its features, like autonomy (i.e. instant identification and response to an invading pathogen with no guidance from the brain) and self-learning, seem to be desired in every (future) security mechanism. However, even at the level of complexity and sophistication that the human immune system has achieved, one cannot claim its 100% reliability. The recent surge of the Ebola virus in Africa with a death rate of approximately 50% has once again proven its susceptibility. This happened despite the couple of centuries people have dedicated to trying to push the limits of immunity by developing and applying vaccines. In other words, people exploit the self-learning capability of the immune system, to manually strengthen it, by injecting weakened pathogens to trigger the generation of antibodies. Ebola escaped the known pattern, as its origin is probably animal-related and appearances among humans have been extremely rare so far, and thus its eradication has never been attempted. Overall, despite the fact that the immune system is never 100% effective, it can achieve high reliability for limited spans of time.

Hence, when projected onto cybersecurity, we may state that: *(i)* no practical security solution can ever be 100% reliable and *(ii)* some security solutions can be highly effective but rather for a short period of time. The following examples from cybersecurity illustrate these claims. An example for *(i)* is SPAM, which has been an issue (a more or less significant one) to communication networks for more than 20 years now. Over all these years, even though there are some solutions that can limit the problem, SPAM has persisted, adapted and spanned to new "territories" e.g. social media. An illustration of *(ii)* is DES (Data Encryption Standard) encryption algorithm developed in the early 1970s at IBM. It was approved as a federal standard in November 1976 and was recommended and utilized till 2001. However, the rapid development of CPU power led to the point where in July 2012, security researchers David Hulton and Moxie Marlinspike were able to recover the DES encryption keys using brute force in less than 24 hours [6]. Therefore, all the confidential data encrypted with DES, even if it was considered secure 20 years ago, is no longer so. So, let us accept that no practical, persistent and perfect security solution exists. Of course, some security measures, like the OTP (One-Time Pad) cipher were mathematically proved as unbreakable. However, they turned out to be impractical and hard to deploy in real-life systems and networks.

The other lessons from nature for cybersecurity are that: *(i)* a security expert should never underestimate a potential threat – some researchers even go a step further and conclude that even overestimating the risk is a good strategy in many circumstances [14]; *(ii)* one should react as soon as the threat emerges, not when it escalates – the reaction time is crucial; *(iii)* threats evolve at least at the pace of security solutions if not faster; *(iv)* the threat may be effectively mitigated but often the solution is short-lived. The failure to contain the Ebola virus in Africa before it proliferated and threatened to become a worldwide epidemic is an example of how the abovementioned rules were not taken into account. If local authorities of the countries where the disease was first diagnosed, as well as the WHO, had not underestimated the risks and reacted together decidedly, the impact of Ebola and the number of deceased would be limited. Moreover, the first infections by this virus were noted back in 1976 and there was ample time to work on a vaccine that could effectively defend humankind in the future. However now, as the virus spread, the risk has amplified as the virus might mutate and cause even more severe damage. A potential cure would then be even harder to create, not to mention the soaring cost of combating the epidemic.

We argue that (future) cybersecurity solutions should take advantage of nature, hence be designed, developed and deployed in a way that covers as many features from the proposed PROTECTION framework as possible. This in turn can potentially yield products that are adaptation-ready, extensible and effective. Noting that current cyber threats possess traits from the discussed framework (as proved earlier), it becomes obvious that so should security systems. Moreover, we believe that, among the presented features in proposed framework, the biggest potential for cybersecurity lies in cooperation and redundancy. Until recently the cooperation feature has been significantly undervalued, while redundancy is unjustly considered for security as a waste of resources. Putting appropriate stress on these two features should open some new paths. A good example of a successful implementation of the cooperation component of the PROTECTION framework (besides presented in Case Study Section) is the system of industry consortiums called ISACs (Information Sharing and Analysis Centers) [9] organized by the US Department of Homeland Security. ISACs federate cybersecurity defense information and distribute it to all of its members and they have fostered much cooperation between security communities within each of the given industries – there is e.g. Financial Services ISAC (FS-ISAC) or the Defense Industrial Base ISAC (DIB-ISAC). This is evidence that cooperation within the white hat community is improving, and soon may be quite effective in mitigating the global threats.